# Single charge control of localized excitons in heterostructures with ferroelectric thin films and two-dimensional transition metal dichalcogenides

Danjie Dai [ab], Xinyan Wang [ab], Jingnan Yang [c], Jianchen Dang [ab], Yu Yuan [ab], Bowen Fu [c], Xin Xie [ab], Longlong Yang [ab], Shan Xiao [ab], Shushu Shi [ab], Sai Yan [ab], Rui Zhu [ab], Zhanchun Zuo [ab], Can Wang *[abd], Kuijuan Jin [abd], Qihuang Gong [c], Xiulai Xu *[c]

[a] Beijing National Laboratory for Condensed Matter Physics, Institute of Physics, Chinese Academy of Sciences, Beijing 100190, China
[b] CAS Center for Excellence in Topological Quantum Computation and School of Physical Sciences, University of Chinese Academy of Sciences, Beijing 100049, China
[c] State Key Laboratory for Mesoscopic Physics and Frontiers Science Center for Nano-optoelectronics, School of Physics, Peking University, 100871 Beijing, China
[d] Songshan Lake Materials Laboratory, Dongguan, Guangdong 523808, China

* xlxu@pku.edu.cn
*canwang@iphy.ac.cn

## Abstract

Single charge control of localized excitons (LXs) in two-dimensional transition metal dichalcogenides (TMDCs) is crucial for potential applications in quantum information processing and storage. However, traditional electrostatic doping method with applying metallic gates onto TMDCs may cause the inhomogeneous charge distribution, optical quench, and energy loss. Here, by locally controlling the ferroelectric polarization of the ferroelectric thin film $BiFeO_3$ (BFO) with a scanning probe, we can deterministically manipulate the doping type of monolayer $WSe_2$ to achieve the p-type and n-type doping. This nonvolatile approach can maintain the doping type and hold the localized excitonic charges for a long time without applied voltage. Our work demonstrated that ferroelectric polarization of BFO can control the charges of LXs effectively. Neutral and charged LXs have been observed in different ferroelectric polarization regions, confirmed by magnetic optical measurement. Highly circular polarization degree about 90 % of the photon emission from these quantum emitters have been achieved in high magnetic fields. Controlling single charge of LXs in a non-volatile way shows a great potential for deterministic photon emission with desired charge states for photonic long-term memory.

## Introduction

High-quality quantum light sources are highly desired to implement quantum key distribution and linear optical quantum computation for information processing. [1-5] Solid-state quantum emitters have attracted more attention because of their addressability and integrability. Many solid-state single-photon sources are based on confined low dimensional materials, including nitrogen vacancies in diamond and semiconductor quantum dots. [6-9] Recently, nonclassical quantum emissions have been observed in semiconductor transition metal dichalcogenides (TMDCs), which were ascribed to the LXs introduced by the point defects or electronic perturbations in the monolayer TMDCs.[6,10-15] By intentionally introducing local strains, these quantum emitters can be site-controlled and arrayed.[16-18] Quantum emitters in tungsten-based TMDCs systems have been investigated in many studies.[6,12,19] For example in $WSe_2$, quantum emitters have been demonstrated from the intervalley defect excitons associated with the hybridization of dark excitons by point-like strain perturbations or vacancies.[20-22]

Since atomically thin layers of TMDCs can be stacked vertically using a simple mechanical exfoliation method, quantum emitters can be integrated into a range of functional heterostructure devices, which makes the electrical charge control possible.[23,24] Recently, the LXs have been studied in the vertically assembled heterostructures using gated electrostatic doping.[25,26] The electric field has been used to tune the band energy of quantum emitters by Stark effect, and to control the fine structure splitting (FSS) by exciton wave function modulation.[27,28] Normally, neutral LXs at zero magnetic field possess FSS with orthogonal linear polarizations caused by electron-hole exchange interactions. In contrast, charged LXs do not have FSS due to the disappearance of exchange interactions by the excess electrons or holes.[29,30] Therefore, deterministic control the charge state of LXs is essential to realize desired exciton recombination.

Typically, the charge density of TMDCs can be controlled by gate-bias tuning.[29,31,32] In this way, a transverse p-n junction has been demonstrated with controlling the gate voltage. However, the metal gate would result in an inhomogeneous charge distribution and direct contact between metal and the monolayer TMDCs would also lead to an optical quench. When the gate voltage is removed, the charge density cannot be maintained. To overcome these, integrating ferroelectric films (FE) with TMDCs provide an strategy on controlling the charge of the LXs, since the spontaneous polarization of FE that can be reversed with external stimuli, and can keep the ferroelectric polarization for a long time.[33-35] In addition, the TMDCs/FE heterostructure largely eliminates interface problems introduced by electrode contacts.[36,37] By using scanning probe technique, ferroelectric domains with good retention can be formed with TMDCs in a non-volatile manner, allowing the design of devices not limited by physical source, drain and gate electrodes.[36] The integrated heterostructures could also provide a way for the seamless integration of data processing and storage.[38]

In this work, a 10% Zn-doped BFO ferroelectric thin film and a monolayer $WSe_2$ were used to build a gate-free heterostructure WSe$_2$/BFO, with which the carrier doping of WSe$_2$ can be controlled by the ferroelectric polarization of BFO. We demonstrate that the carrier doping of WSe$_2$ in the downward polarization (P$_{down}$) region is p-type, while in the upward polarization (P$_{up}$) region it is n-type at room temperature. At low temperature without magnetic field, the photoluminescence (PL) peaks of LXs are mostly doublets at the domain wall, while the singlets are mainly located in the P$_{down}$ and P$_{up}$ regions. The lack of FSS is a feature of charged LXs,[29,30] while neutral LXs are more likely to be produced at the domain wall. The presence of charged and neutral LXs is further confirmed by magneto-optical measurements. Highly circular polarization of PL of about 90 % from quantum emitters is observed under a high magnetic field. Using electrical polarization of BFO controlling the single charges of LXs provides a method for achieving quantum emitters with desired charging states in a non-volatile way.

## Results and discussion

### Scanning probe microscopy

A schematic plot of a WSe$_2$/BFO heterostructure is shown in Fig. 1(a), and the fabrication process is summarized in the section of Methods. The BFO film has been grown on 20 nm strontium ruthenate (SRO) buffered (001) strontium titanate (STO) substrates. Mechanical exfoliation and dry transfer were used to create the van der Waals interface of WSe$_2$/BFO. Optical microscopic images of the monolayer WSe$_2$ before and after the transfer are shown in Supplementary (Figure S1). The atomic force microscopy (AFM) image is shown in Fig.1 (b). The orange line in Fig. 1(e) shows that the thickness of WSe$_2$ is about 1 nm, indicating a monolayer, which is also consistent with Raman and PL results at room temperature (as presented in Supplementary Figure S2). The ferroelectric polarization of BFO can be reversed by applying a voltage.[34] The piezoelectric force microscopy (PFM) image of the WSe$_2$/BFO heterostructure are presented in Fig.1(c). The Zn-doping BFO film exhibits distinct downward polarization (P$_{down}$) self-poling polarization in white box region and perfect polarization-voltage (P-V) loop with a large remanent polarization value ~60 μC cm$^{-2}$ (as shown in Fig. 1(d)). In order to obtain an upward polarization (P$_{up}$) region, we applied a -15 V voltage at the metal probe to reverse the downward polarization of the BFO film. As shown by the green line in Fig. 1(e), the out of plane PFM phase of BFO is different by about 180° between the P$_{up}$ and P$_{down}$ regions, confirming that the polarization direction can be effectively switched by the external electric field. The upper surface of the BFO is rich with positive charges in the P$_{up}$ region, and the opposite is true in the P$_{down}$ region. The junction between P$_{down}$ and P$_{up}$ is the domain wall, as labelled P$_{dw}$. The domain wall is at the boundary of positive and negative charges, so it is possible to form a neutral environment. The geometry of the monolayer WSe$_2$ in the PFM image matches that of the AFM image, indicating that the monolayer WSe$_2$ was not damaged during polarization reversal with applying voltage.

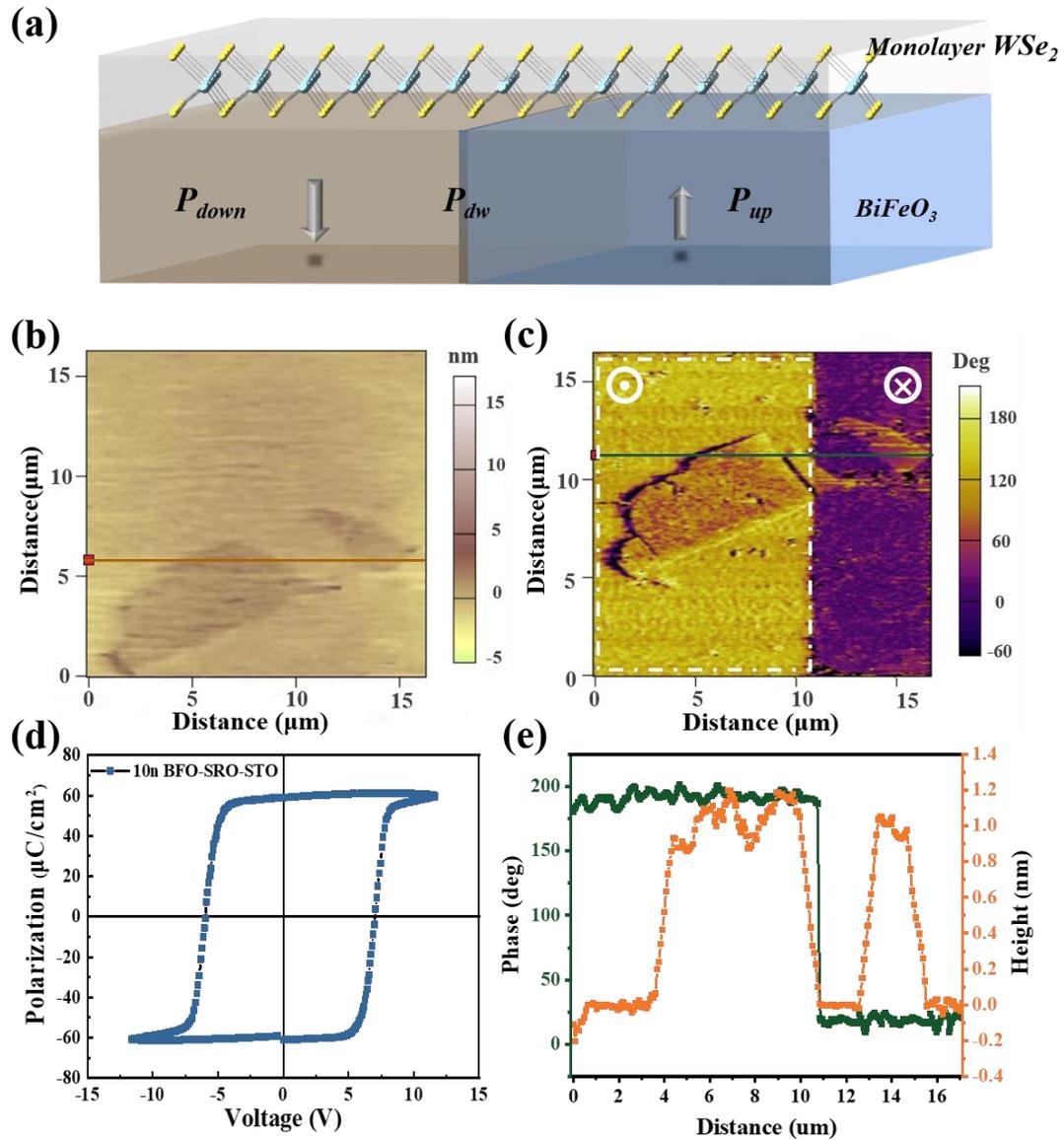

**Fig. 1** The ferroelectric properties of BFO and the scanning probe microscopic images of WSe$_2$/BFO. (a) A schematic plot of the WSe$_2$/BFO heterostructure. (b and c) AFM and PFM images of WSe$_2$/BFO. The yellow region represents the P$_{up}$ region, and the violet region represents the P$_{down}$ region. (d) Electric P–V hysteresis loops. The BFO film has a dramatic ferroelectric switch, and its remanent polarization is about 60 μC cm$^{-2}$. (e) The thickness of monolayer WSe$_2$ measured along the solid orange line in (b) and out-of-plane PFM phase of BFO along the green solid line in (c). The thickness of monolayer WSe$_2$ is approximately 1 nm. The out-of-plane PFM phase of BFO is different by about 180° in the P$_{up}$ and P$_{down}$ regions.

Photoluminescence spectra at zero magnetic field. To elucidate the role of the ferroelectric nature of BFO on the charge density of monolayer WSe$_2$, the temperature-dependent PL measurements have been performed from 35 K to 273 K, as shown in Fig. 2(a). Because of the large binding energy and direct band gap of monolayer TMDCs, direct band recombination with different type of carrier doping can be resolved by PL spectroscopy even at room temperature.[39,40] Different from the intrinsic exciton ($X_{in}$), the localized exciton is quickly quenched as the temperature rises. The emission of 1.73 eV at 35 K corresponds to the neutral exciton $X_0$. The energy difference between the neutral and charged excitons allows us to determine the carrier doping type of the device. In the $P_{up}$ region, two different trions states of negatively charged exciton ($X_{in}^-$) are located at 1.688 eV and 1.696 eV under 35 K, which correspond to the triplet trion ($X_{inT}^-$) and the singlet trion ($X_{inS}^-$). While in the $P_{down}$ region, the positively charged exciton Xin+ is singlet state.[41-44] The emission of negatively charged intrinsic exciton $X_{in}^-$ in the $P_{up}$ region is red-shifted by about 10 meV relative to the $X_{in}^+$ in the $P_{down}$ region at low temperature. The PL from $P_{up}$ region also undergoes a red shift of about 15 meV at 273 K, which is consistent with previous reports on PL with metal electrostatic gating.[41-44] This demonstrates that the direct band gap recombination of WSe$_2$ switched from positively charged excitons to negatively charged excitons from the natural downward polarization $P_{down}$ region to the polarization flipped $P_{up}$ region. The $g$-factors of intrinsic excitons in the $P_{down}$ and $P_{up}$ regions are extracted with polarization-resolved magneto-PL at 4.2 K (as shown in Supplemental Figure S3). The $g$-factors in the two regions are both close to 4, further confirming that they are direct band gap emission.[12,42,44-47]

Figure 2(b-d) depict the PL spectra of monolayer WSe$_2$ at different positions in the three regions of $P_{down}$, $P_{up}$, and $P_{dw}$ at 4.2 K. As shown in Fig. 1c, the WSe$_2$ flake is approximately 3 μm at the junction, which is large enough to avoid lateral confinement effects. Many narrow g-fs were observed in the low-energy region, and the full widths at half maximum of these peaks fall between 100 and 500 μeV. We attribute these peaks to the localized single exciton transitions induced by the interaction or combination of crystal defects and local strain during the transfer process. Compared with the $X_{in}$ the LXs have a narrower linewidth and lower energy. Intriguingly, the PL peaks of these LXs exhibit different characteristics at the zero magnetic field, some of them are doublets and others are only single peaks (singlets). The PL peaks of LXs located at the domain wall are mainly doublets and the two peaks have the cross-linearly polarized emission (as shown in Supplemental Figure S4), which indicates that the doublets at the domain wall originate from the same localized exciton. The splitting is the result of FSS caused by anisotropic electron-hole exchange interactions in asymmetric confinement potentials.[48,49] However, in our experiments, singlets are mainly observed in the $P_{down}$ and $P_{up}$ regions. Due to the ferroelectric polarization of BFO, the density of negative charges in monolayer WSe$_2$ increases in the $P_{up}$ region, making it easier to produce negatively charged excitons (X$^-$), while positively charged excitons (X$^+$) in the $P_{down}$ region. The charged exciton at its lowest energy state consists of two holes (electrons) with opposite spins and one electron (hole).[26,29,50,51]

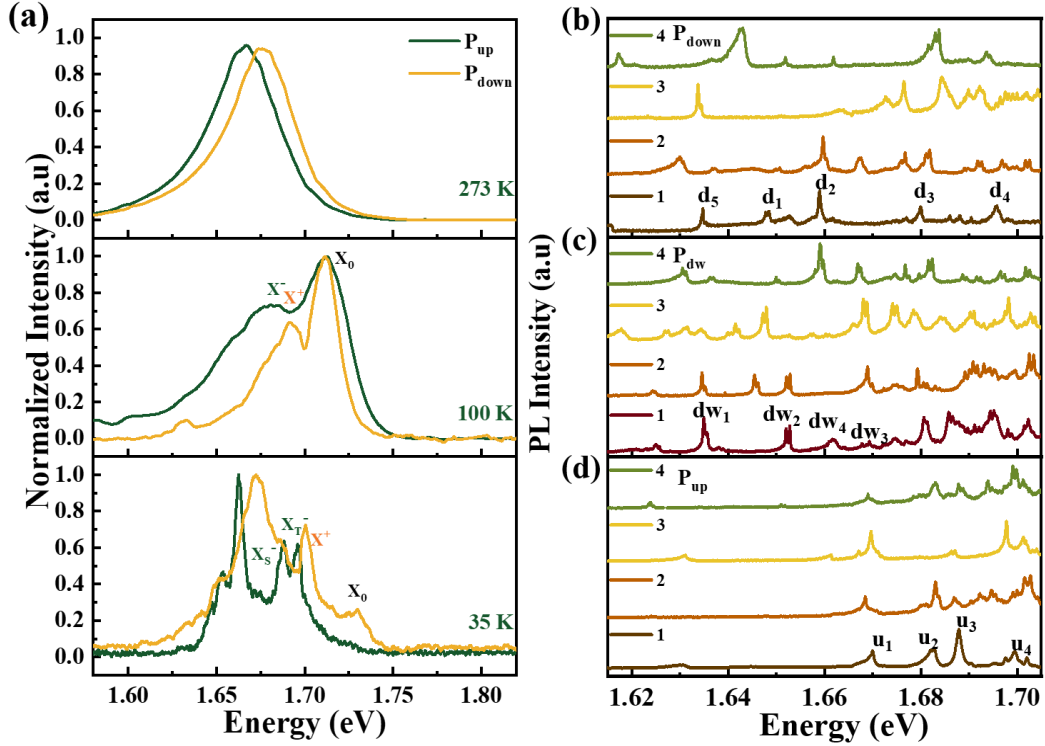

**Fig. 2** PL spectra at zero magnetic field. (a) Temperature-dependent PL spectra from 35 K to 273 K in the $P_{down}$ and $P_{up}$ regions. (b–d) PL spectra under different positions of the $P_{down}$ region (a), the $P_{dw}$ region (b), and the $P_{up}$ region (c) at 4.2 K.

**Magneto photoluminescence spectra measurements**

To further investigate magnetic optical properties of LXs, the magneto-PL measurements were performed in three regions of $P_{down}$, $P_{up}$, and $P_{dw}$. Fig. 3 shows the magnetic response of the labeled PL peaks of LXs in Fig. 2 (b-d). The typical X-shaped dispersion can be observed in all three regions. The FSSs at zero magnetic field of the doublets in all three regions are approximately 700-800 μeV, which are consistent with the previously reported values.[52,53] The emission of the LXs at zero magnetic field is linearly polarized. This is because the monolayer WSe$_2$ valley degrees of freedom and spin degrees of freedom are locked, then the eigenstates at zero magnetic field are a linear superposition of the left-hand circularly polarized excitonic state $|K>$ and the right-hand circularly polarized excitonic state $|K'>$, and they have equal weights.[54]

$$|X_L> = \frac{1}{\sqrt{2}}(|K> - |K'>) \quad (1)$$

$$|X_U> = \frac{1}{\sqrt{2}}(|K> + |K'>) \quad (2)$$

As the magnetic field increases, the degeneracy is gradually lifted.

$$|X_L> = N_L\left(|K> + \left(\frac{g\mu_B B}{\delta} - \sqrt{1 + \left(\frac{g\mu_B B}{\delta}\right)^2}\right)|K'>\right) \quad (3)$$

$$|X_U> = N_U\left(|K> + \left(\frac{g\mu_B B}{\delta} + \sqrt{1 + \left(\frac{g\mu_B B}{\delta}\right)^2}\right)|K'>\right) \quad (4)$$

$$E_L = E_0 - \frac{1}{2}\sqrt{(g\mu_B B)^2 + \delta^2} \tag{5}$$

$$E_U = E_0 + \frac{1}{2}\sqrt{(g\mu_B B)^2 + \delta^2} \tag{6}$$

where $L$ refers to the low energy peak, while $U$ denotes the high energy peak, $\mu_B$ is the Boltzmann constant, $g$ represents the Landé factor, and $\delta$ denotes the FSS at zero field, $N_L$ and $N_U$ are magnetic field dependent normalization constants. The energy splitting ($\Delta$) between high energy peak and low energy peak as a function of applied magnetic field is shown in Fig. 3(d-f). For doublets, we extract the exciton $g$-factor by using the expression.

$$\Delta = \sqrt{(g\mu_B B)^2 + \delta^2} \tag{7}$$

In contrast to the behaviors of doublets at low magnetic fields, in which the energy splitting increases quadratically with the magnetic field, the energy splitting of the singlets in $P_{down}$ and $P_{up}$ is a linear function of magnetic field due to the absence of zero field FSS.

$$\Delta = \sqrt{(g\mu_B B)^2} \tag{8}$$

The $g$-factors of singlets (peak $d_2$, $d_3$, and $d_4$) in the $P_{down}$ region are 9.93, 9.84, and 9.97, and those in the $P_{up}$ region (peak $u_1$ and $u_3$) are 10.43 and 10.27, respectively. The $g$-factors of all the doublets in all the three regions are approximately 9.30. It is consistent with the previous reports on the $g$-factor for LXs of monolayer $WSe_2$.[10,51] This is because ferroelectric polarization affects the charge level of monolayer $WSe_2$, leading to the production of positively charged LXs in the $P_{down}$ region and negatively charged LXs in the $P_{up}$ region. On the other hand, carrier distribution induced by the domain wall provides more opportunities to generate neutral LXs.

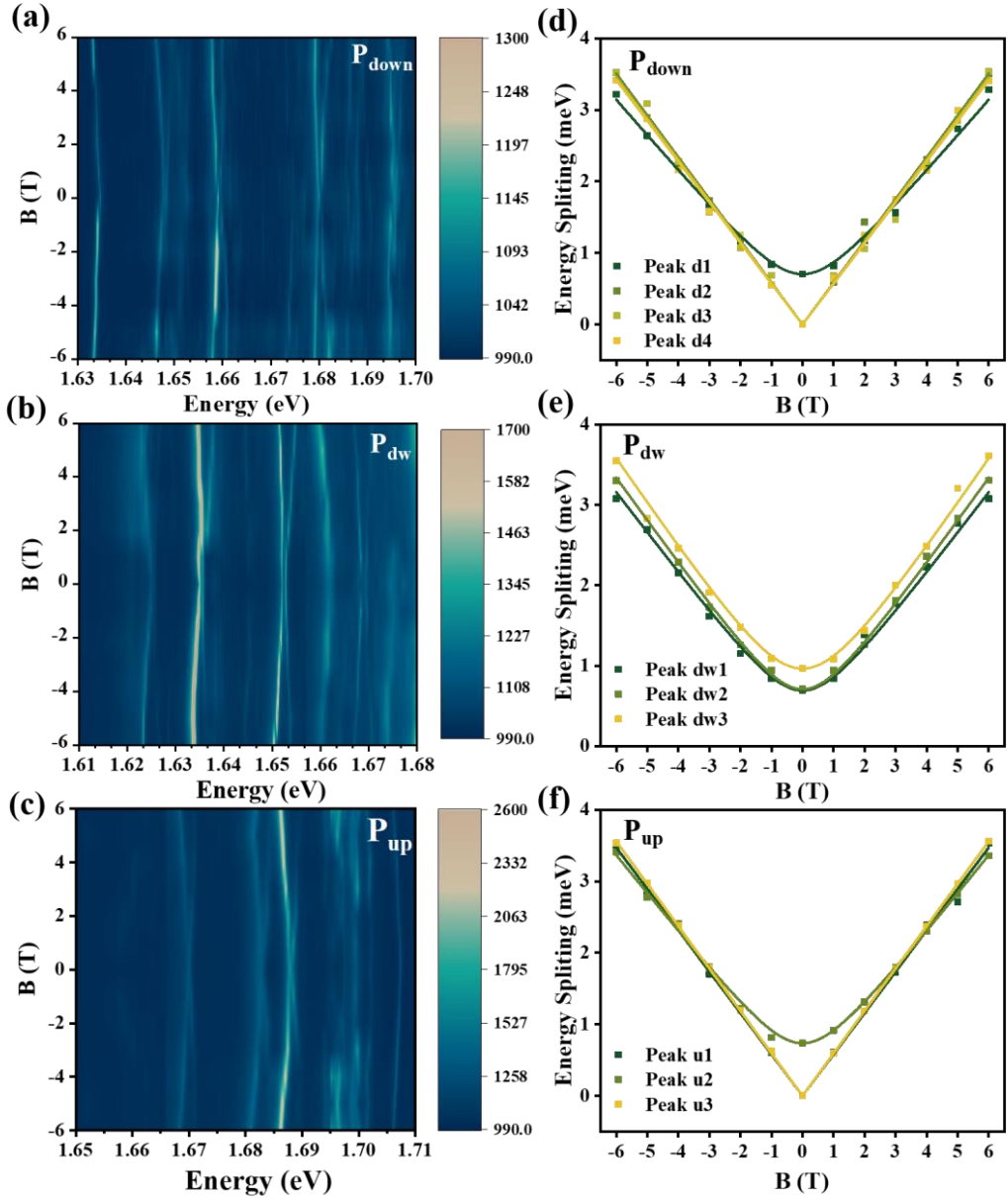

**Fig. 3** Magneto-PL in P$_{down}$, P$_{up}$, and P$_{dw}$ regions by applying a Faraday configuration magnetic field (perpendicular to the sample). Contour plots of PL spectra as a function of magnetic field in different regions, P$_{down}$ (a), P$_{dw}$ (b), and P$_{up}$ (c). (d–f) Energy splitting as a function of the applied magnetic field. The solid lines represent the fitting of experimental data. Peak d$_2$, d$_3$, and d$_4$ in the P$_{down}$ (d) region are singlets with *g*-factors of 9.93, 9.84, and 9.97, respectively. Peak d$_1$ is a doublet with *g*-factor and FSS values of 9.01 and 703.1 μeV, respectively. Peak dw$_1$, dw$_2$, and dw$_3$ in the P$_{dw}$ region (e) have *g*-factors and FSSs of 9.02, 9.41, 9.65 and 692.01, 708.96, and 882.02 μeV, respectively. The *g*-factors of peak u$_1$, u$_2$, and u$_3$ in the P$_{up}$ region (f) are 10.22, 10.43, and 10.27, respectively, and the FSS of peak u$_2$ is 735.02 μeV.

Unlike the PL peaks of LXs in Fig. 3 with distinct Zeeman splitting, we also found several PL peaks whose high energy branch component is strongly suppressed over the entire magnetic field range. As displayed in Fig. 4(a–c), the PL peak shifts into the low energy direction as the magnetic field increases, while the high energy branch component is unresolved. Fig. 4(d–f) illustrate the variation of the energy with increasing magnetic field. For peak $dw_4$ in the $P_{dw}$ region, the variation curve of energy with magnetic field is parabolic, whereas the peak $d_5$ in $P_{down}$ and peak $u_4$ in $P_{up}$ are straight lines. The solid line represents the fitting by eqn (5), from which we can extract the $g$-factor and zero field FSS. The $g$-factors of peak $d_5$, peak $dw_4$, and peak $u_4$ are 8.76, 10.60, and 7.65, respectively. The zero field FSS values cannot be extracted for peak $d_5$ and peak $u_4$ in $P_{down}$ and $P_{up}$ regions. The lack of zero field FSS is the hallmark of charged LXs. This matches well with the results we discussed earlier that the charged LXs were generated in the $P_{down}$ and $P_{up}$ regions via the charge control of $WSe_2$ by the ferroelectric polarization of BFO. On the other hand, for peak $dw_4$ in the $P_{dw}$ region, a zero field FSS of 497.23 μeV can be obtained, which is a little smaller than the reported value of $WSe_2$.[10,51] The missing of the high energy branch in the magnetic spectrum can be explained by two reasons. One is thermal relaxation, with which carriers preferably occupy the lower energy levels.[6,26,55,56] Second, strong asymmetry might induce the high energy branch with a small optical oscillator strength.[10,54] In our experiment, the strong asymmetry of the confining potential might be induced by the unevenness of the substrate or wrinkling during the transfer process

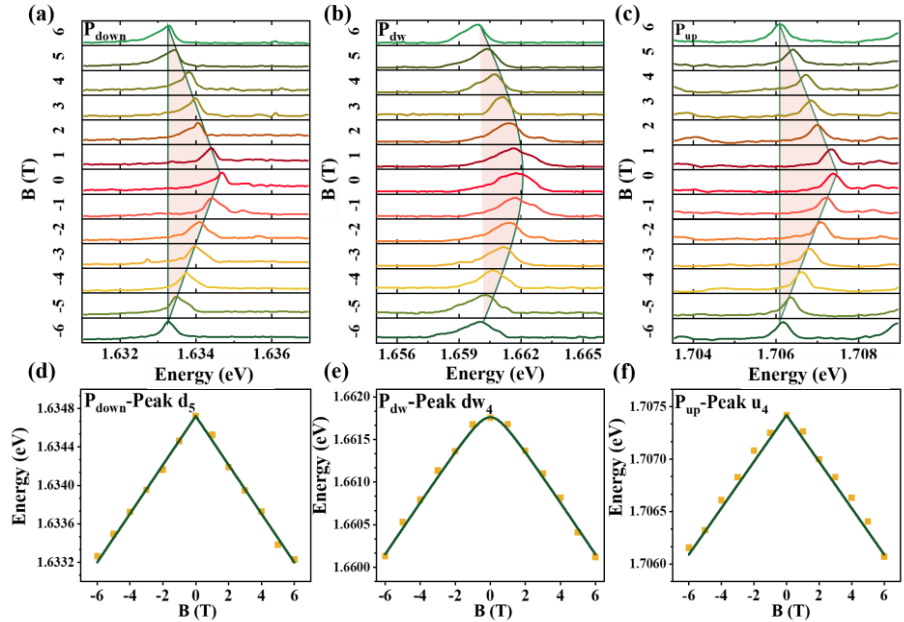

**Fig. 4** Magneto-PL of peak $d_5$, $dw_4$, and $u_4$ by applying a Faraday configuration magnetic field. (a–c) Magneto-PL spectra of the peak $d_5$ (a), peak $dw_4$ (b), and peak $u_4$ (c). (d–f) Energy as a function of the applied magnetic field. The solid lines represent the data fitted by eqn (5) and (6). g-factor of peak $d_5$ in the $P_{down}$ region is 8.76 (d), peak $u_4$ in the $P_{up}$ region is 7.65 (f). For peak $dw_4$ in $P_{dw}$ region, the g factor of 10.60 and FSS of 497.23 μeV are extracted.

**Polarization-resolved magneto photoluminescence spectra measurements**

According to the eqn (1)–(4), the circular polarization degree of LXs increases with magnetic field increasing. Fig. 5 (a and b) depict the circularly polarized PL spectra of peak $u_4$ with a linearly polarized excitation. The PL intensity of σ− emission is significantly higher than that of σ+ with a positive magnetic field, while σ+ emission is much stronger for a negative magnetic field. Fig. 5(c) illustrates the degree of circular polarization degree extracted from the integral intensity of PL peaks. The degree of circular polarization is defined as

$$P = \frac{I_{\sigma-} - I_{\sigma+}}{I_{\sigma-} + I_{\sigma+}} \tag{9}$$

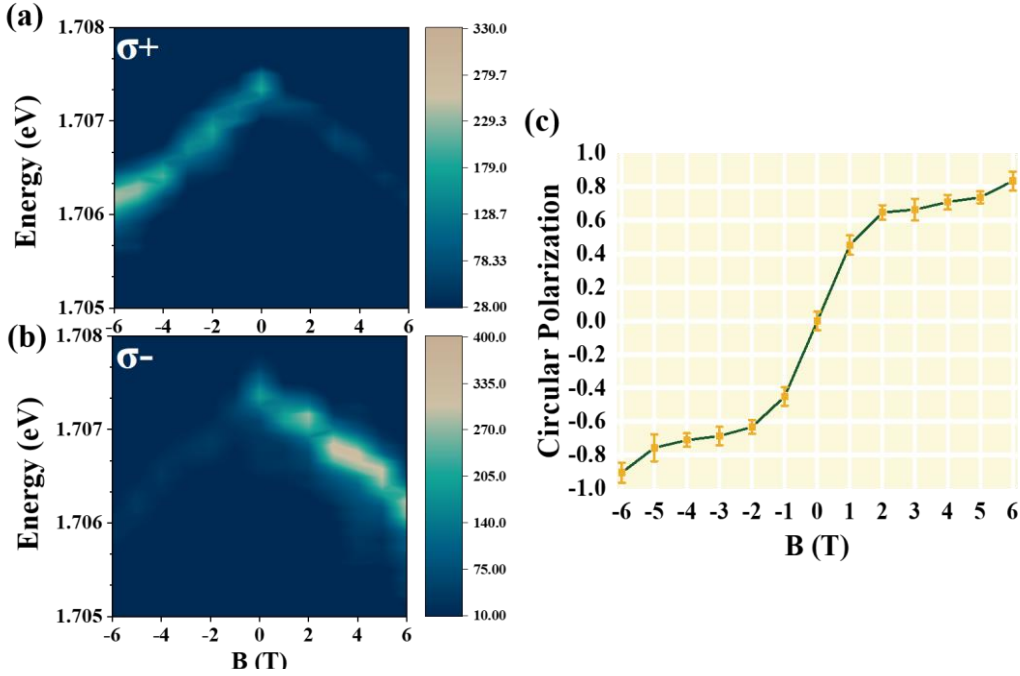

**Fig. 5** Contour plot of polarization-resolved magneto-PL peak $u_4$ in a Faraday configuration magnetic field. (a and b) Polarization-resolved magneto-PL under linearly polarized excitation for $\sigma_+$ (a) and $\sigma_-$ (b) collection. (c) The circular polarization degree of the lower energy branch of peak $u_4$ obtained from (a) and (b).

## Methods

### Device fabrication

To fabricate WSe$_2$/BFO heterojunction, the Bi$_{1.3}$Fe$_{0.9}$Zn$_{0.1}$O$_3$ films were grown firstly on ~20 nm strontium ruthenate (SRO)-buffered (001) strontium titanate (STO) substrates with a laser molecular beam epitaxy system. The WSe$_2$ monolayer films on scotch tape was prepared by the mechanical exfoliation method. The monolayer was then transferred onto the BFO film using PDMS as a transfer medium, resulting in a heterostructure. AFM and PFM characterization was done a commercial AFM system (Asylum Research, MFP-3D). Ti/Pt coated silicon probes with a tip radius of about 28 nm were used for collecting and recording the PFM images.

**PL measurement**

The samples were placed on a three-dimensional piezoelectric stage with a superconducting magnet cryostat, in which a vertical magnetic field of ±9 T and a low temperature of 4.2 K can be provided. All PL measurements were performed using a confocal microscope system. A large numerical aperture objective lens of 0.82 was used to excite the sample and collect PL light. The sample was excited with a continuous wave 532 nm laser. The PL signal was coupled into a single-mode fiber and acquired through a grating spectrometer with a liquid nitrogen-cooled charge-coupled device.

## Conclusion

We have demonstrated successfully that the ferroelectric polarization of BFO can control the single charges of LXs of monolayer $WSe_2$ with $WSe_2$/BFO heterojunction. Ferroelectric gating can maintain the charges of exciton for a long time in a non-volatile and low-energy manner. The doping type of direct band gap emission of $WSe_2$ changes from n-type to p-type, which was confirmed from the $P_{down}$ region to the $P_{up}$ region with PL peak energy shifting at different temperatures. For LXs, the singlet with the absence of FSS in $P_{down}$ and $P_{up}$ regions is attributed to the extra charge introduced by the ferroelectric polarization. Most neutral LXs have been observed in the $P_{dw}$ region at the domain walls, which is confirmed with magneto-PL measurements. Moreover, a high circular polarization with 90% LXs was obtained at a high magnetic field of 6T. Singly charge state control in LXs in TMDCs with the FE substrate provides a method for long-term single spin detection and spin storage in a non-volatile manner, which could be useful for TMDCs-based information processing. In this work, the domination of zero-field FSS at the domain walls indicates the polarization confinement, which lays the groundwork for the physical exploration of the role of polarization confinement at the domain walls on localized excitons.

## Author contributions

X. Xu, C.W., K. J. and Q. G. conceived and planned the project. D. D and X. W fabricated the devices. D. D., J. Y., J. D., Y. Y., B. F., X. Xie, L. Y., S. X., S. S., S. Y., R. Z. and Z. Z. performed the optical measurement. All authors discussed the results and wrote the manuscript.

## Conflicts of interest

The authors declare no competing interests.

## Acknowledgements

This work was supported by the National Key Research and Development Program of China (Grant No. 2021YFA1400700), the National Natural Science Foundation of China (Grants No. 62025507, 11934019, 11721404, 11874419, 62175254 and 12174437), the Key R&D Program of Guangdong Province (Grant No. 2018B030329001), the Strategic Priority Research Program (Grant No. XDB28000000) of the Chinese Academy of Q3 Sciences.